\documentstyle[12pt,epsfig]{article}
\begin{document}

\begin{minipage}{14cm}
\vskip 2cm
\end{minipage}
\vskip 2cm

\begin{center}
{\bf $Z$-SCALING OF JET PRODUCTION AT TEVATRON }


\vskip 5mm

{M.V. Tokarev$^{\star}$} and T.G. Dedovich

\vskip 5mm

{\small
 {\it
Veksler and Baldin Laboratory of High Energies,\\
Joint Institute for Nuclear Research,\\
141980, Dubna, Moscow region, Russia}
\\
$^\star${\it
E-mail: tokarev@sunhe.jinr.ru
}
}
\end{center}

\vskip 5mm

\begin{center}
\begin{minipage}{150mm}
\centerline{\bf Abstract}
New data on jet production obtained by the CDF and D0 Collaborations
at the  Tevatron in Run II are analyzed in the framework of $z$-scaling.
Properties of data $z$-presentation are discussed.
Physics interpretation of the scaling function $\psi(z)$ as
a probability density to produce
a particle with the formation length $z$ is argued.
It was shown that these experimental data confirm $z$-scaling.
\end{minipage}
\end{center}
\vskip 1cm

\vspace*{2cm}
\begin{center}
{\it
 Submitted to the XVII International Seminar on High Energy Physics\\
 "Relativistic Nuclear Physics  and Quantum Chromodynamics",\\
September 27 - October 2, 2004, Dubna, Russia }
\end{center}

\newpage

{\section{Introduction}}
The production of very large transverse momentum hadron jets
in hadron-hadron  collision at high energies at SpS in CERN
observed  by UA2 and UA1 Collaborations confirming
the hints of jet production in the experiment of AFS Collaboration at ISR
was an convincing experimental proof of quarks and gluons existence
confirming the theory of strong interaction  - Quantum ChromoDynamics (QCD).
The wide study of the phenomena is performed in $\bar p-p$ collisions
at highest energy at the Tevatron \cite{Korytov,Rimondi}.

Jets are experimentally observed as a strong correlated group
of particles in space-time which are copiously produced
at hadron colliders. They are result of hard scattering
of quarks and gluons and their subsequent transformation
into real particles.

New era of QCD precision measurements is starting at hadron colliders RHIC,
Tevatron and LHC.
The study of the energy and angular dependencies of jet
and dijet cross sections, invariant mass distribution of jets,
structure and content of jets and their fragmentation properties
is considered to give enough information both for verification of
the theory (QCD and SM) and search for new physics phenomena
in new kinematical domain.

Fragmentation of partons into hadrons is one of the least
understanding  feature of QCD.
Even though the primary scattering process is described in term
of perturbative QCD the hadronization chain contains very low, respective
to the parent parton, $p_{\bot}$ hadrons. Therefore the whole  process
is clearly a non-perturbative  phenomena involving final state interactions which
have to conserve color and baryon number.
The quarks and gluons carry color charge and are essentially massless
in the theoretical calculations. A  hadron jet has no color charge and
often large invariant mass. Thus jets are ambiguous objects and should be treated
in such a way that these unavoidable ambiguities
do not play an important role \cite{S.Ellis}.

A search for general properties of jet production
in hadron-hadron collisions is of great interest,
especially  in  connection with  commissioning
such large accelerators of nuclei and proton as the RHIC at BNL
and  the LHC at CERN.
The main physical goals of the
investigations at these colliders are to search for and study
Quark Gluon Plasma (QGP) - the hot and extremely dense phase of the nuclear
matter, Higgs boson and particles of new generation predicted
by supersymmetry theories, and to understand origin of the proton spin.

Jets are traditionally  considered  to be one of the good probes for
study the hard interaction between quarks and gluons in
the surrounding nuclear matter and search for indication on phase transition.
Jet production in collisions of polarized protons at the RHIC give a new tool
for study of enigmatical Nature of particle spin as well.

In the report we present the results of our analysis
of new data on jet production obtained by the CDF and D0 Collaborations
at the Tevatron in Run II in the framework of $z$-scaling concept \cite{Zscal,Dedovich,zppg}.
The concept is based on the fundamental principles  such as self-similarity,
locality and fractality of structure of colliding objects, interaction of their
constituents and mechanism of particle formation.

The scaling function $\psi(z)$ and scaling variable $z$ used for presentation
of experimental data are expressed via the observable quantities,
such as an inclusive cross section
and multiplicity density. The function $\psi$ has simple physical
interpretation as the probability density to form a jet with formation length $z$.

We would like to emphasize that the properties of $z$-presentation
for jet production give evidence on fractality of jet formation
mechanism at very small scale up to $ 10^{-4}$~Fm. This is the
region where the fractal geometry of space-time
\cite{Nottale,Mandelbrot,Imr} itself could play an important role
for search for general regularities of all fundamental interactions.

The results of our new analysis are found to be in good agreement
with previous ones \cite{Dedovich}  and are considered as a new
confirmation of $z$-scaling at the Tevatron.


\vskip 0.5cm
{\section{Z-scaling }}
In the section we would like to remind some basic ideas and definitions
of $z$-scaling. As shown in \cite{Zscal,Dedovich,zppg} self-similarity
of high-$p_T$ particle formation reveals itself as possibility
to describe physical process in terms of the scaling function $\psi(z)$
and scaling variable $z$. The function is expressed  via the invariant cross section
$Ed^3\sigma/dp^3$  and the multiplicity density $dN/d\eta$ as follows
\begin{equation}
\psi(z) = -{ { \pi s} \over { (dN/d\eta) \sigma_{in}} } J^{-1} E { {d^3\sigma} \over {dp^3}  }
\label{eq:r7}
\end{equation}
Here, $s$ is the center-of-mass collision energy squared, $\sigma_{in}$ is the
inelastic cross section, $J$ is the corresponding Jacobian.
The factor $J$ is the known function of the
kinematic variables, the momenta and masses of the colliding and produced particles.

The normalization equation
\begin{equation}
\int_{0}^{\infty} \psi(z) dz = 1
\label{eq:r8}
\end{equation}
allows us to interpret the function $\psi(z)$ as a probability density to produce
a particle with the corresponding value of the variable $z$.

The variable $z$ as established in \cite{Zscal,Dedovich,zppg}
 reveals property of fractal measure. It can be written in the form
\begin{equation}
z = z_0 \Omega^{-1}.
\label{eq:r9}
\end{equation}
The finite part $z_0$ is the ratio of the transverse
energy released in the binary collision of constituents
and the average multiplicity density $dN/d\eta|_{\eta=0}$.
The divergent part
$\Omega^{-1}$ describes the resolution at which the collision of
the constituents can be singled out of this process.
The quantity $\Omega(x_1,x_2) = m(1-x_{1})^{\delta_1}(1-x_{2})^{\delta_2}$
represents relative number of all initial
configurations containing the constituents which carry fractions
$x_1$ and $x_2$ of the incoming momenta.
The $\delta_1$ and $\delta_2$ are the anomalous fractal
dimensions of the colliding objects (hadrons or nuclei).
The momentum fractions $x_1$ and $x_2$ are determined in a way to
minimize the resolution $\Omega^{-1}(x_1,x_2)$ of the fractal
measure $z$ with respect to all possible sub-processes
satisfying the momentum conservation law.
The variable $z$ is interpreted  as a particle formation length.

The general regularities of $z$-scaling for jet production in
$p-p$ and $\bar p-p$ collisions at the ISR, Sp$\bar{\rm p}$S and
Tevatron were established in \cite{Dedovich}. In the present
report the regularities are verified by using new experimental
data obtained by the D0 and CDF Collaboration at the Tevatron in
Run II.

\vskip 0.5cm {\section{Jets at hadron colliders}} A jet is
experimental observed as a strong correlated group of particles in
space-time. At low collision energies high-$p_T$ hadrons was only
observed and considered as result of hard scattering of elementary
hadron constituents. At high collision energy  $\sqrt s $ jets are
copiously produced at hadron colliders such as Sp$\bar{\rm p}$S
and Tevatron. They are considered as an experimental signature of
quarks and gluons interactions.

Figure 1 demonstrates the high-$p_T$ spectra of $\pi^0$-mesons
produced at the ISR and azimuthal correlations of jets produced in
$\bar p-p$ collisions  at the  Sp$\bar{\rm p}$S and Tevatron and
in $p-p$ collisions at the RHIC.

{\subsection {Jet definition}}
In interaction of colliding hadrons two (or more)
highly  collimated collection of particles having approximately equal
transverse momentum are observed. These collimated beams of particles in space-time
are called jets. The strong correlation of high-$p_T$ particles from the jets
in the azimuthal plane is one of main features of jet production.
The high-$p_T$ hadrons are considered to be produced by hadronization of
quarks and gluons.
A tipical dijet event is assumed to consist of hard interaction
and underlying event.
The last one includes initial and final gluon and quark radiation,
secondary semi-hard interactions, interaction between remnants,
hadronization and jet formation. Thus the procedure for extraction information on
hard constituent interactions and comparison with theoretical calculation
in the framework of QCD is an indirect one and sophisticated algorithms
of data analysis are required.

{\subsection{Cone algorithm}}
A standard definition of a jet to facilitate
comparisons of measurements from different
experiments and with theoretical predictions
was accepted in the Snowmass Accord \cite{ConeJet}.

The Snowmass Jet Algorithm defines a jet as a collection of partons,
particles or calorimeter cells contained within a cone of an open angle $R$.
All objects in an event have a distance from the jet center
$\Delta R_i=\sqrt {  {(\phi_i-\phi_{jet})}^2 + {(\eta_i-\eta_{jet})}^2    }    $,
where $(\phi_{jet}, \eta_{jet})$ define  direction of the jet and
$(\phi_{i}, \eta_{i})$ are the coordinates of the parton, particle
or center of the calorimeter cell. If $\Delta R_i \le R$ then the object is a part of jet.
The transverse energy $E_T$ and direction of jet are defined by formulas
$$E_T = {\sum_{i\in R_i \le R}} E_T^i$$
\begin{equation}
\eta_{jet} = (1 /E_T) \cdot {\sum_{i\in R_i \le R}}\eta^i \cdot E_T^i
\label{eq:r11}
\end{equation}
$$\phi_{jet} = (1 /E_T) \cdot {\sum_{i\in R_i \le R}}\phi^i \cdot E_T^i $$
An iterative procedure for finding the jets given by the Snowmass algorithm
includes determination of jet seeds, jet cone formation, determination
of the transverse energy and direction of jet. The definition of jet seed
is not given by the algorithm. The Snowmass Accord does not deal with
jet overlap.

At the parton level seeds could be partons, points lying between pairs
of partons, a set of points randomly located in the $\eta-\phi$ space.
Experimentally the seed could be defined as any cell in calorimeter or
clusters of calorimeter cells.
Therefore there are different treatment of jets at the parton
and calorimeter level.
To accommodate the difference between the jet definitions
at the parton and calorimeter level
in the modified Snowmass algorithm  a purely phenomenological
parameter  $R_{sep}$  has been suggested \cite{Kunszt}.
At the parton level  $R_{sep}$  is the distance between two partons
when the clustering algorithm switches from a one jet  to a
two jet final state, even though both partons are contained within
the jet defining cone.
The maximum allowed distance $\Delta R$ between two partons
in a parton jet divided by the cone size used: $R_{sep}= \Delta R/R$.
The value of $R_{sep}$ depends on details of the jet algorithm used
and the experimental jet splitting and merging scheme.

{\subsection{Clustering $k_T$-algorithm}}
Clustering algorithms in contrast to cone algorithms,
which globally find the jet direction, successively
merge pairs of nearby vectors. The order
in which vectors are recombined into jets defines the algorithm.
The $k_T$-algorithm  combines vectors based on their relative
 transverse momentum.

Several variants of the clustering $k_T$-algorithm for hadron collisions
have been proposed \cite{kTJet}.
It is designed to be independent of the order in which the seeds are processed.
It is infrared and collinear safe.
The initial seeds are all charged particles with $k_{T,i}$ in a given $\eta$-range.
Each  seed is labeled as prejet.  Measure or closeness criterion in phase space
is defined for each prejet and pair of prejets as follows
\begin{equation}
 d_i  = k_{T,i}^2
\label{eq:r12}
\end{equation}
\begin{equation}
d_{i,j}  = min \{ k_{T,i}^2, k_{T,j}^2 \} \cdot \Delta R_{i,j}^2 /R^2
\label{eq:r13}
\end{equation}
Here $R$  is a jet cone size, $\Delta R_{i,j}$ is the distance between
 two prejets (i and j)
 in $(\eta, \phi)$ space.
The procedure of jet finding includes the next steps:
computation of the measure of all prejets and all pairs of prejets;
finding  of the prejet or pair of prejets with the smallest measure $d_{min}$;
promotion it to a "jet" and remove its particles from considerations
if $d_{min}$ arises from a single prejet; combination the pair into a new prejet
and recompute measure for all prejets and pairs of prejets if $d_{min}$
arises from a pair of prejets; continuation previous steps until all prejets
have been promoted to jet.

Here we would like to note that different modifications
of cone and clustering algorithms have been used in analysis
of experimental and theoretical jet data to obtain their compatibility
\cite{ConeJet,kTJet,Abbott,CDFjet}.
It is especially important for study of soft and hard processes
contribute to jet formation \cite{Stuart}.
Transformation of quarks and gluons into real particles
is considered to include evolution of constituent structure and their interactions
at different scales. It corresponds to different scheme used for evolution of
parton distribution functions \cite{DGLAP,BFKL,CCFM}.
Therefore general regularities which can be extracted from experimental data
could give complementary constraints on theoretical models of jet formation
and new insight on origin of particle mass as well.

\vskip 0.5cm
{\section{$Z$-scaling and jet production at Tevatron in Run II}}

In this section we analyze new data on inclusive cross section
of jet production in $\bar p-p$ collisions at $\sqrt s = 1960$~GeV
obtained by the D0\cite{Begel,Chleb}
and CDF \cite{Meyer,Tonnes}  Collaborations at the Tevatron
in Run II and compare them with our previous results \cite{Dedovich}.

{\subsection{Energy independence of $\psi$ }}
The production  hadron jets at the Tevatron probes the highest
momentum transfer region currently accessible  and thus
potentially sensitive  to a wide  variety of new physics.
The information on inclusive jet cross section at high
transverse momentum range is the basis to test QCD,
in particular to extract the strong coupling constant $\alpha_S(Q^2)$,
the parton distribution functions and to constrain
uncertainties for gluon distribution in the high-$x$ range.
In Run II, as  mention in \cite{Tonnes}, the measurement of jet
production  and the sensitivity to new physics will profit
from the large integrated luminosity  and the higher cross
section, which is associated with  the increase in the center-of-mass
energy from 1800 to 1960~GeV.
Therefore the test of $z$-scaling for jet production in
$\bar p-p$ collisions in new kinematic range is of great interest
to verify scaling features established in our previous analysis \cite{Dedovich}.

The D0 and CDF Collaborations have carried out the
measurements \cite{Begel,Meyer} of transverse spectra
of single inclusive cross sections of jet production
at $\sqrt s = 1960~GeV$. In the D0 \cite{Begel}
and  CDF \cite{Meyer} experiments single jets
were registered in the $0.0<|\eta|<0.5$ and $0.1<|\eta|<0.7$ ranges,
respectively. The data were used in present analysis.

New data on inclusive cross sections of jet production in $\bar p-p $
collisions obtained by the D0 Collaboration at the Tevatron
in Run II are presented in Figure 2(a) \cite{Begel}.
Spectrum of jet production is measured at $\sqrt s =1960$~GeV in
the pseudorapidity  and  transverse momentum
ranges  $|\eta|<0.5$ and  $p_T = 60-560$~GeV/c, respectively.
Data $p_T$- and $z$-presentations  are shown in Figure 2(b) and 2(c), respectively.
Note that results of present analysis of new D0 data are in a good agreement
with our results \cite{Zscal} based on the data  \cite{Abbott}
obtained by the same Collaboration in Run I.
The energy independence and the power law (the dashed line in Figure 2(c))
of the scaling function $\psi(z)$ are found to be as well.

The dependence of single  jet cross section on transverse momentum of
jet in $\bar p-p$ collisions at $\sqrt s = 630, 1800$ and 1960~GeV
is shown in Figure 3(a). The data \cite{CDFjet,Meyer}
covers momentum range $p_T=10-560$~GeV/c.
The energy dependence of jet cross section
is observed to be strong from $\sqrt s = 630$ to 1800~GeV.
and weak from 1800 to 1960~GeV.
Data $p_T$ and $z$-presentation is shown in Figure 3(b) and 3(c), respectively.
 As seen from Figure 3(c)
new data \cite{Meyer} are in agreement with other ones obtained in Run I.
The energy independence of $\psi(z)$ is observed up to $z \simeq 4000$.
Asymptotic behavior of scaling function is described by the power
law, $\psi(z)\sim z^{-\beta}$ (the dashed line in Figure 3(c)).
The slope parameter $\beta$ is energy independent over a wide $p_T$-range.

{\subsection{Angular independence of $\psi$}}
Let us consider the angular dependence of $p_T$- and $z$-presentations
new of D0 \cite{Chleb} and CDF \cite{Tonnes} data.
The D0 and CDF collaborations have carried out the
measurements \cite{Chleb,Tonnes} of the angular dependence
of the single inclusive cross sections of jet production
at $\sqrt s = 1960~GeV$. In the D0 experiment \cite{Chleb}
a single jet was registered in the range $0.0<|\eta|<2.4$.
In the CDF experiment \cite{Tonnes} jets were registered in the ranges
$0.1<|\eta_1|<2.8$.

We would like to note that the strong dependence of the cross
sections on the angle of produced jet was experimentally found
at the SpS and the Tevatron in Run I.

Figures 4(a) and 5(a) show the
dependence  of the cross sections
of the $\overline p+p \rightarrow jet+X$ process on the transverse
momentum $p_{\bot}$ at $\sqrt s=1960$~GeV for different rapidity
intervals, $0.5<|\eta|<2.4$ and $0.1<|\eta|<2.8$, measured by the
D0 and CDF Collaborations, respectively. The $p_T$-presentation of
new data \cite{Chleb,Tonnes}  demonstrates the strong angular dependence as well.
The qualitative regularities of jet spectra at $\sqrt s =1960$~GeV are
similar to ones at $\sqrt s =630$~GeV and 1800~GeV.

We verify the hypothesis of the angular scaling for
$z$-presentation of the data for jet production in $\bar p-p$ collisions.
The angular scaling of data $z$-presentation means
that the scaling function $\psi(z)$ at given energy $\sqrt s$
has the same shape over a wide $p_T$  and
pseudorapidity range of produced jets.

Figure 4(b,c) and 5(b,c) demonstrate $p_T$ and $z$-presentation of the
D0 \cite{Chleb} and CDF \cite{Tonnes} data sets, respectively.
Taking into account errors of the experimental data  we can conclude that
the  data confirm the angular scaling of $\psi(z)$.
Nevertheless it is necessary to note that some points (last five and seven points
corresponding to the (1.4,2.1) and (2.1,2.8) pseudorapidity ranges of the data \cite{Tonnes}
and two last points corresponding to the (1.5,2.0) and (2.0,2.4)  pseudorapidity ranges
of the data \cite{Chleb}) deviate from the power law.
This is not real indication of $z$-scaling violation.
The reason of the deviation is impossibility to take correctly into account the
kinematical conditions of constituent subprocess  due to large pseudorapidy bins for
reconstructed jets. The smaller angular binning and more higher statistics of data
are necessary to resolve the problem.

{\subsection{Jet multiplicity density $dN/d\eta$}}
The important ingredient of $z$-scaling is the multiplicity density
$ \rho(s,\eta) \equiv  dN/d\eta   $.
The quantity is well determined in analysis of high-$p_T$ particle production.
The energy dependence of $\rho(s)$ for charged hadrons produced in $p-p$
and $\bar p-p$ collisions  at $\eta=0$ is measured up to $\sqrt s =1800$~GeV.
The dependence is used to construct $z$ and $\psi$ for different pieces of particles.

In the case of jet production the quantity is not well determined.
It is connected with the experimental and theoretical determination of jets.
In the analysis \cite{Dedovich} the normalized jet multiplicity density
$\rho(s)/\rho_0$ has been used.
The value  of $\rho(s)/\rho_0$ at the normalization point  $\sqrt s =1800$~GeV
is equal to 1.

Figure 6 shows the dependence of $\rho(s)/\rho_0$ on the collision energy $\sqrt s$.
At the accessible energy range the dependence is well described by the power law shown
in Figure 6  by the dashed line.

\vskip 0.5cm
{\section{Conclusion}}
Analysis of new experimental data on jet production in $\bar p-p$ collisions
obtained at the Tevatron in Run II by the CDF and D0 collaborations
in the framework of data $z$-presentation  was performed.

The scaling function $\psi(z)$ and scaling variable $z$ expressed
via the experimental quantities, the invariant inclusive cross
section $Ed^3\sigma/dp^3$  and the jet multiplicity density
$\rho(s,\eta)$ are constructed.
The scaling function $\psi$ is interpreted  as a
probability density to produce a jet with the formation length $z$.

The general regularities of jet production
(the energy and angular independence of $\psi$,and the power law)
found at the ISR, SpS and  Tevatron in Run I
are confirmed in new kinematical range ($\sqrt s =1960$~GeV and $p_T = 10-550$~GeV/c).
Results of our analysis of new experimental data are found to be
in good agreement with results obtained by the D0 and CDF Collaboration in Run I.
The obtained results are new evidence that mechanism of jet formation
reveals self-similar and fractal properties over a wide transverse momentum range.

Thus we conclude that new data obtained at the Tevatron confirm the general
concept of $z$-scaling. The further inquiry and search for violation
of the scaling could give information on new physics phenomena in high energy
hadron collisions and determine domain of validity of the strong
interaction theory and the Standard Model itself.

\vskip 5mm
{\large \bf Acknowledgments.}
The authors would like to thank I.Zborovsk\'{y}, Yu.Panebratsev,
and O.Rogachevski for collaboration and numerous fruitful
and stimulating discussions of the problem.

\newpage
\begin{minipage}{4cm}

\end{minipage}

\vskip 2cm
\begin{center}
\hspace*{2.5cm}
\parbox{10.cm}{\epsfxsize=10.cm\epsfysize=10.cm\epsfbox[95 95 400 400]
{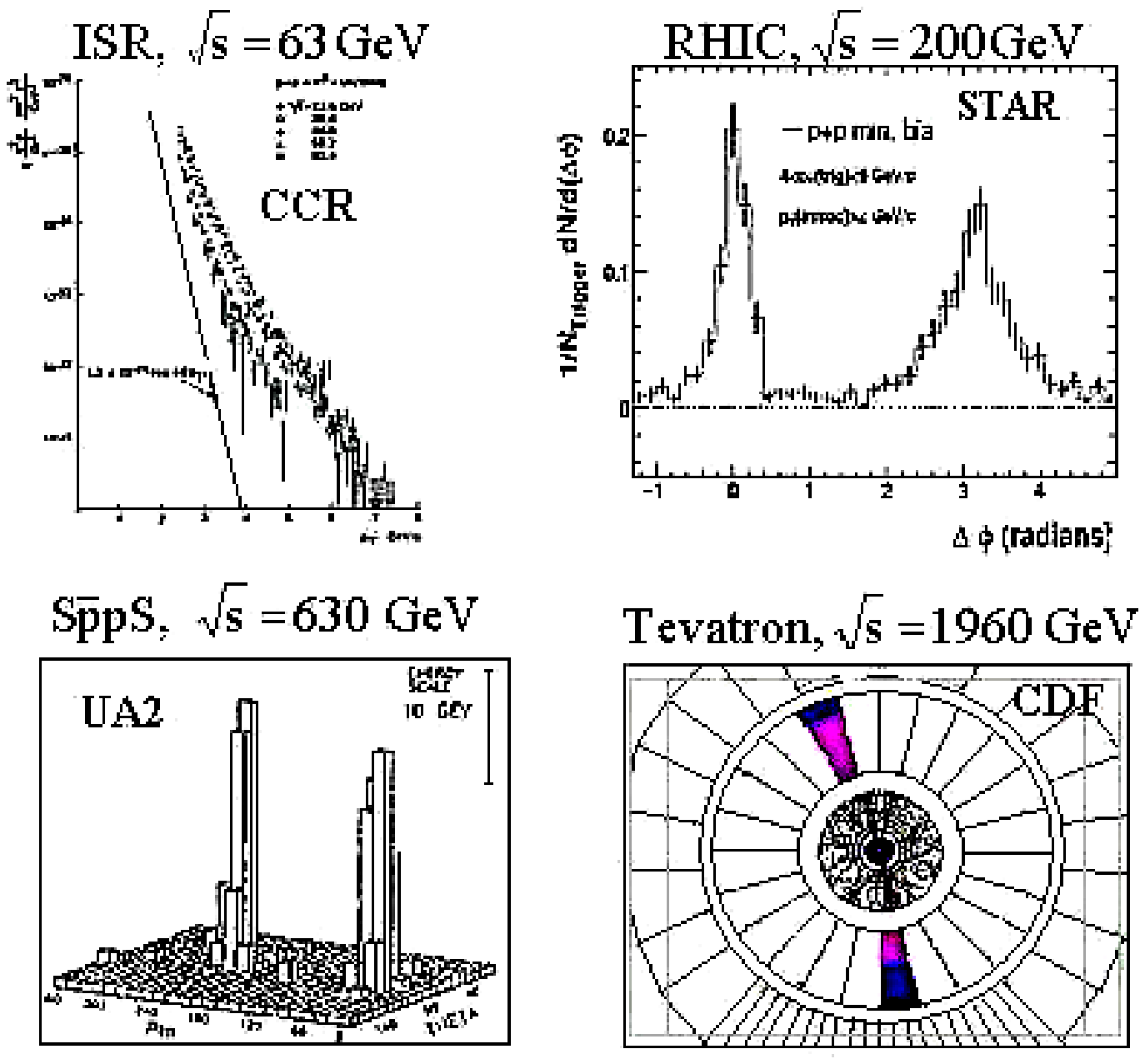}{}}

\vskip 5cm

{\bf Figure 1.} Jets at the ISR, RHIC, Sp$\bar{\rm p}$S  and
Tevatron colliders.
\end{center}

\newpage
\begin{minipage}{4cm}

\end{minipage}

\vskip 2cm
\begin{center}
\hspace*{-1.5cm}
\parbox{4.5cm}{\epsfxsize=4.5cm\epsfysize=4.5cm\epsfbox[95 95 400 400]
{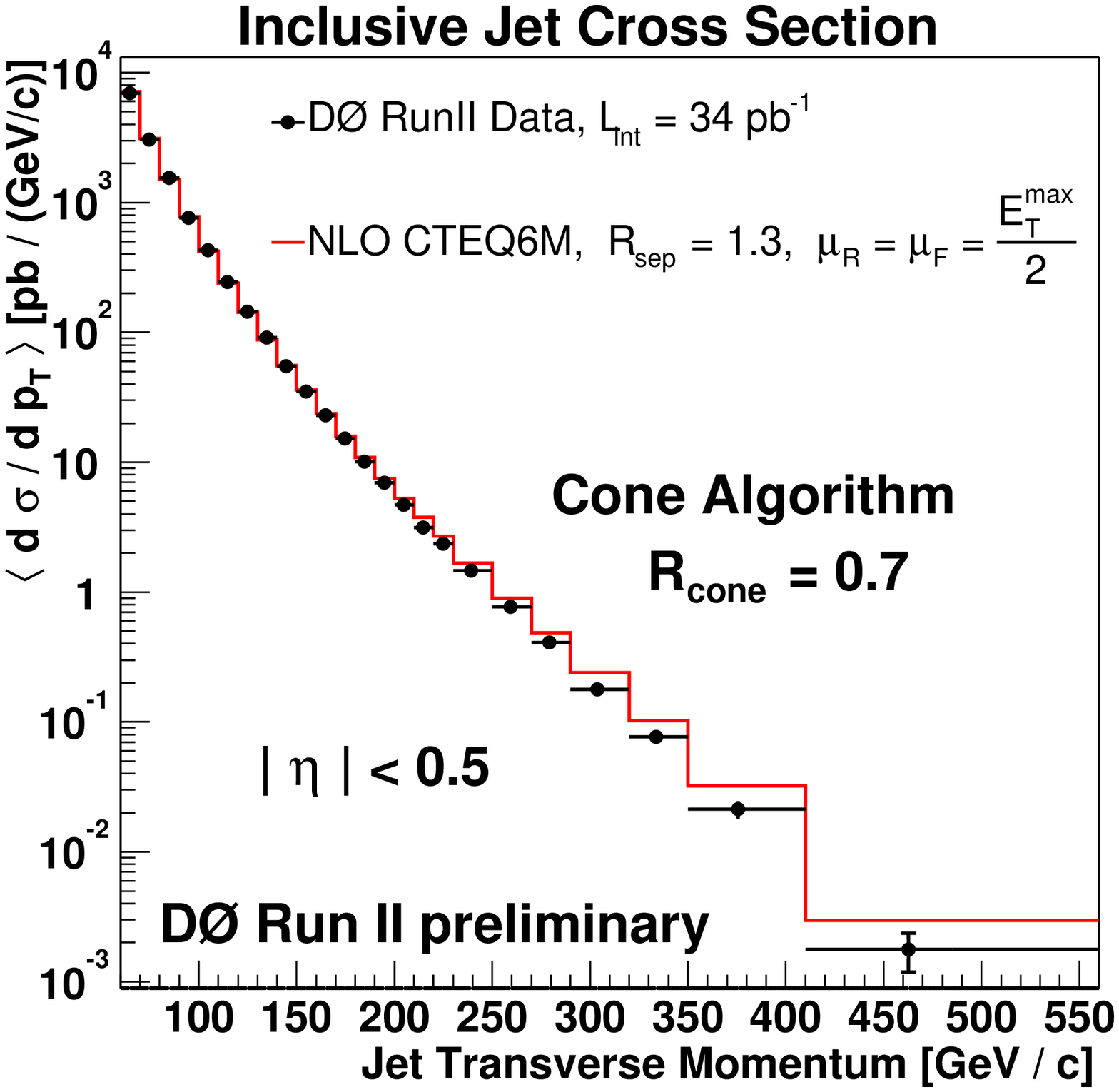}{}}
\vskip 1.5cm
\hspace*{0.cm} a)
\end{center}

\vskip 5cm

\begin{center}
\hspace*{-2.5cm}
\parbox{5cm}{\epsfxsize=5.cm\epsfysize=5.cm\epsfbox[95 95 400 400]
{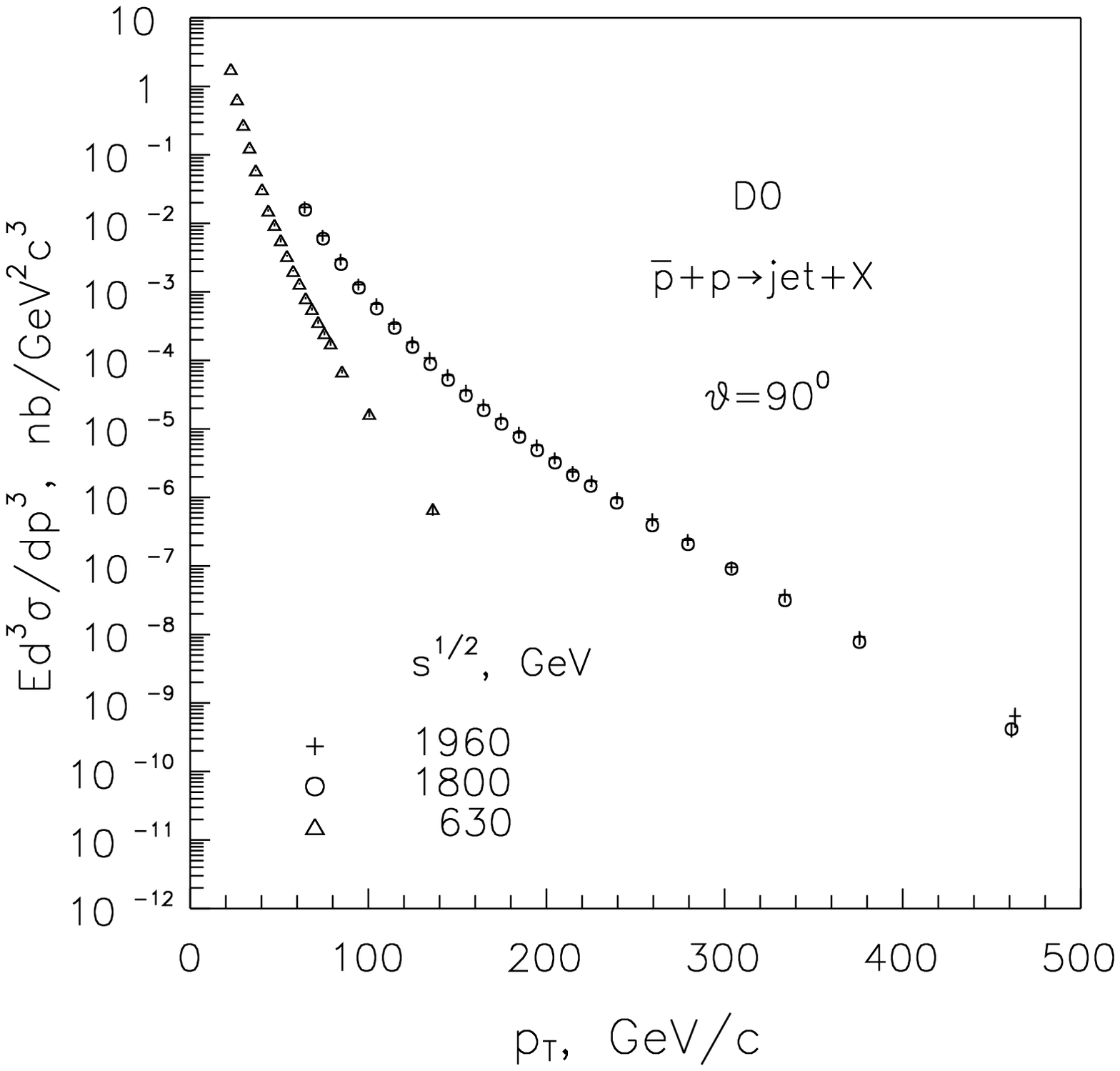}{}}
\hspace*{3cm}
\parbox{5cm}{\epsfxsize=5.cm\epsfysize=5.cm\epsfbox[95 95 400 400]
{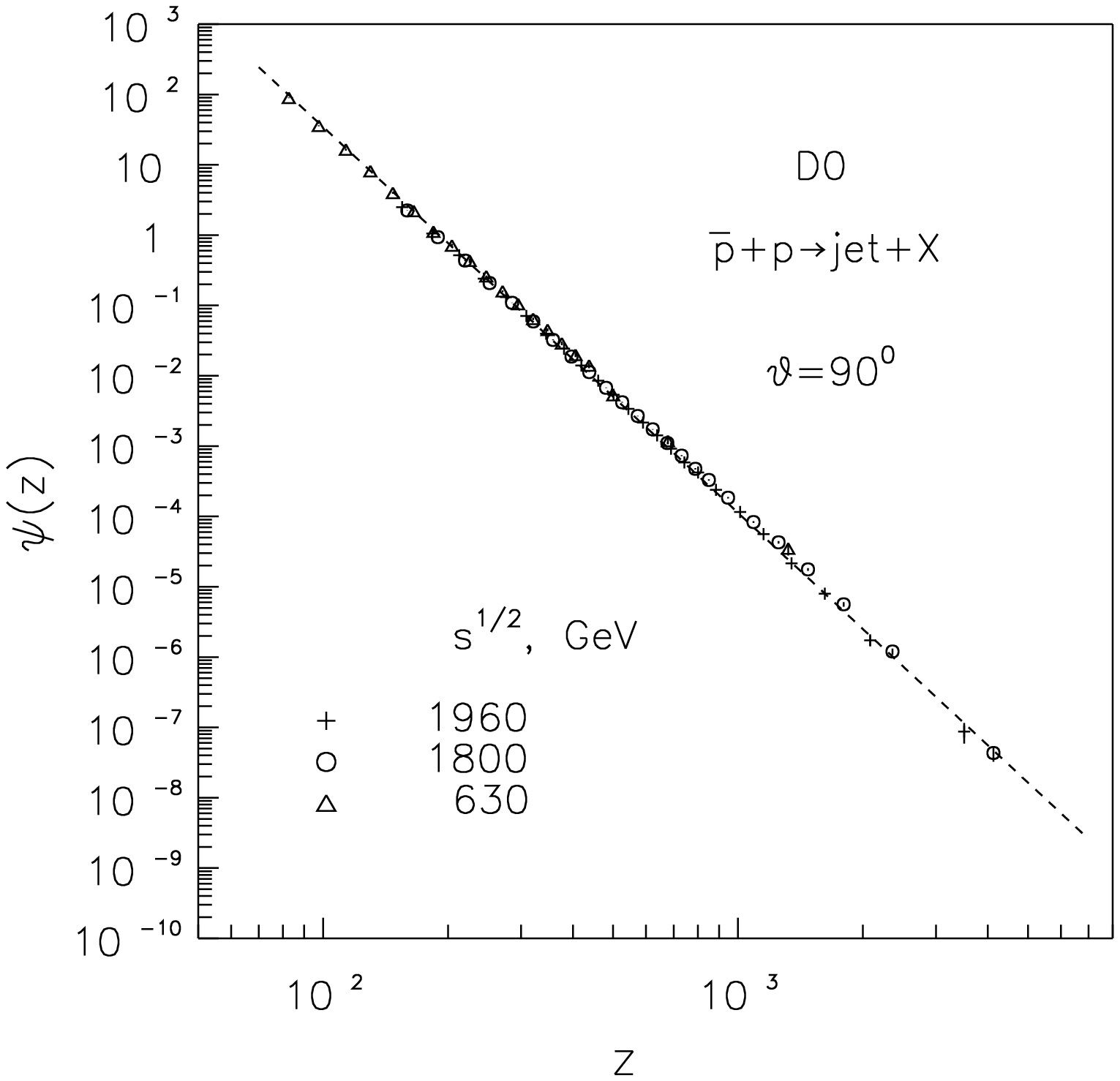}{}}
\vskip -1.cm
\hspace*{0.cm} b)\hspace*{8.cm} c)\\[0.5cm]
\end{center}

{\bf Figure 2.}
(a) The preliminary D0 data  \cite{Begel} on inclusive spectrum $d\sigma /dp_T$
of one jets produced in $\bar p-p $ collisions at $\sqrt s = 1960$~GeV
in the central pseudorapidity range $|\eta|<0.5$ as
a function of the transverse momentum.
(b) The D0 data  on invariant cross section $Ed^3\sigma/dp^3$
of jet production at $\sqrt s =630,1800$~GeV \cite{Abbott}
and  1960~GeV \cite{Begel} in $p_T$- and (c) $z$-presentations.
The dashed line represents the power fit to the data.

\vskip 0.5cm
\newpage
\begin{minipage}{4cm}

\end{minipage}

\vskip -2cm
\begin{center}
\hspace*{-1.5cm}
\parbox{5.5cm}{\epsfxsize=5.5cm\epsfysize=5.5cm\epsfbox[95 95 400 400]
{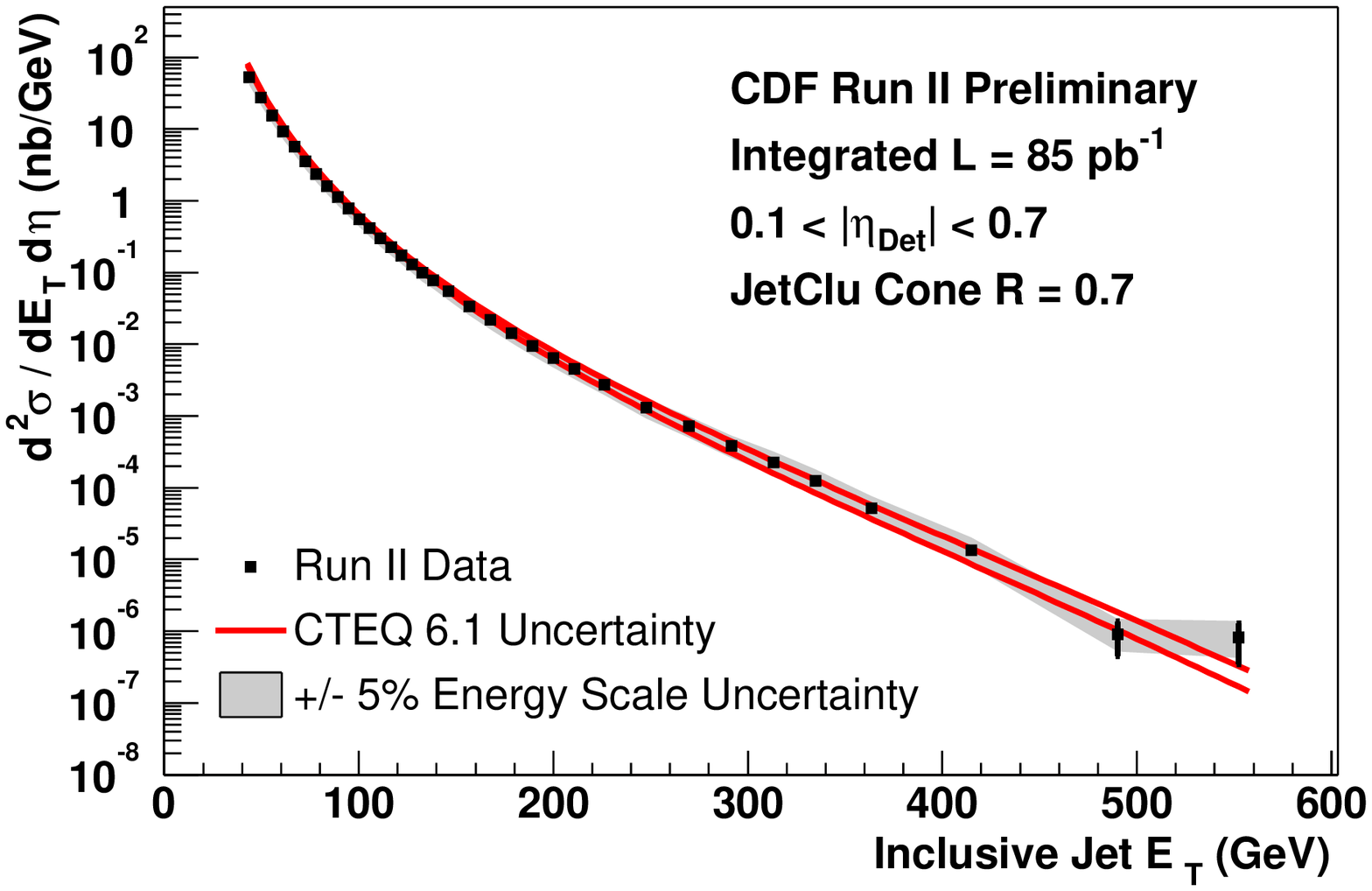}{}}
\vskip 2.5cm
\hspace*{0.cm} a)
\end{center}

\vskip 5cm

\begin{center}
\hspace*{-2.5cm}
\parbox{5cm}{\epsfxsize=5.cm\epsfysize=5.cm\epsfbox[95 95 400 400]
{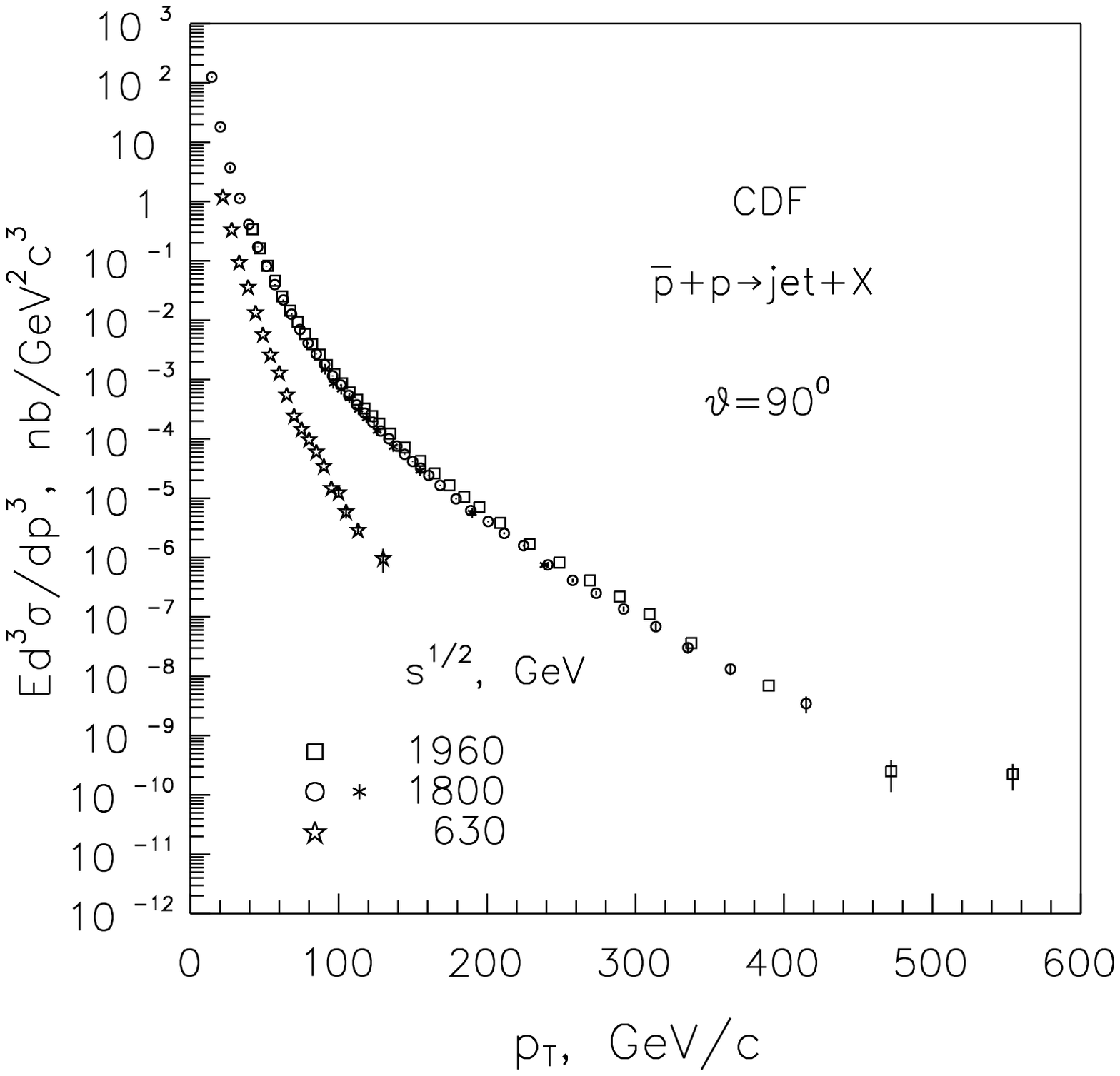}{}}
\hspace*{3cm}
\parbox{5cm}{\epsfxsize=5.cm\epsfysize=5.cm\epsfbox[95 95 400 400]
{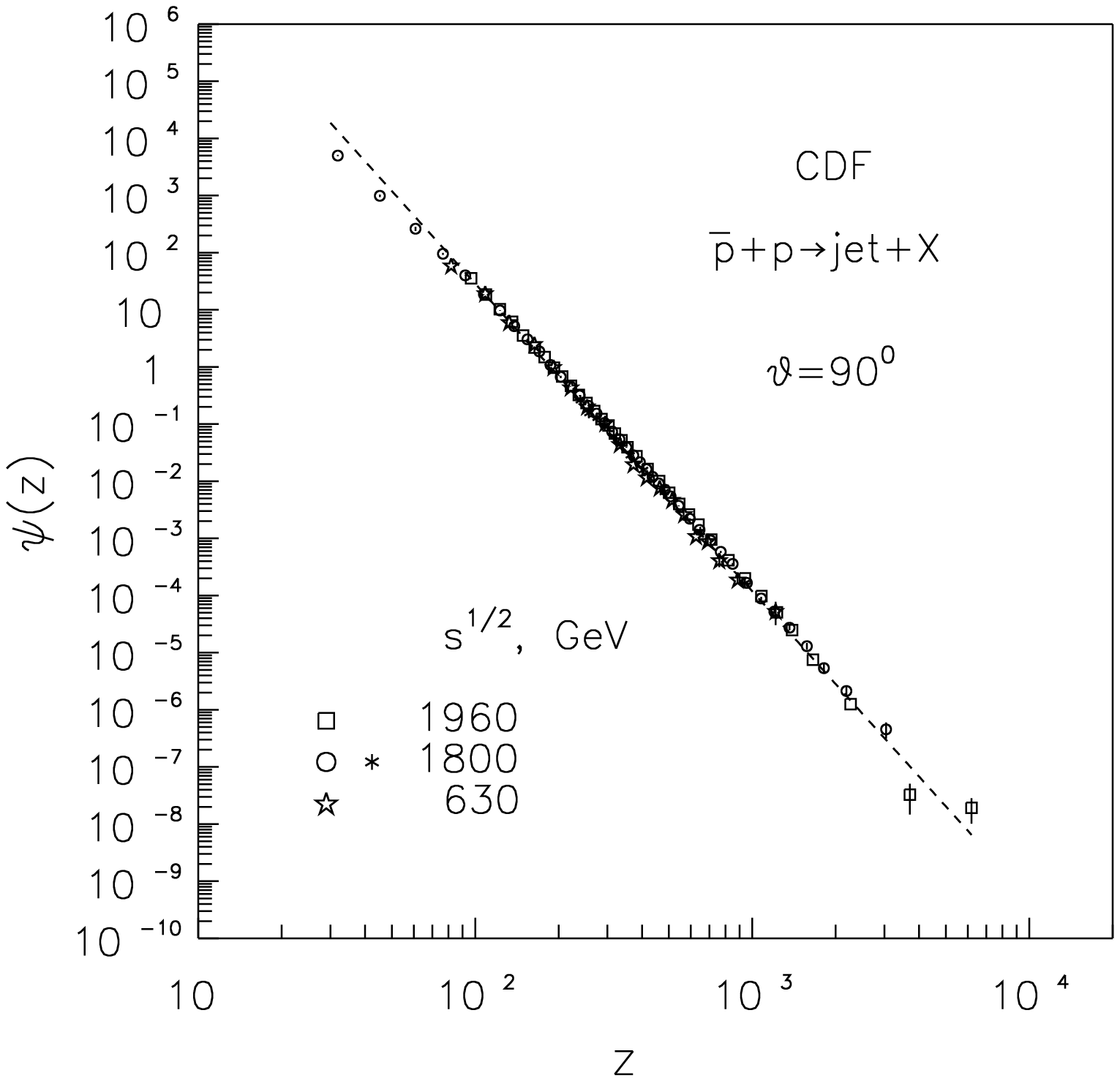}{}}
\vskip -1.cm
\hspace*{0.cm} b)\hspace*{8.cm} c)\\[0.5cm]
\end{center}

{\bf Figure 3.}
a) The preliminary CDF data \cite{Meyer} on inclusive spectrum $d^2\sigma/dE_Td\eta$
of one jets produced in $\bar p-p $ collisions at $\sqrt s = 1960$~GeV
in the central pseudorapidity range $0.1<|\eta|<0.7$ as
a function of the transverse momentum.
(b) The CDF data  on invariant cross section $Ed^3\sigma/dp^3$
of jet production at $\sqrt s =630,1800$~GeV \cite{CDFjet}
and  1960~GeV \cite{Meyer} in $p_T$- and (c) $z$-presentations.
The dashed line represents the power fit to the data.
\vskip 0.5cm

\newpage
\begin{minipage}{4cm}

\end{minipage}

\vskip 2cm
\begin{center}
\hspace*{-1.5cm}
\parbox{5.5cm}{\epsfxsize=5.5cm\epsfysize=5.5cm\epsfbox[95 95 400 400]
{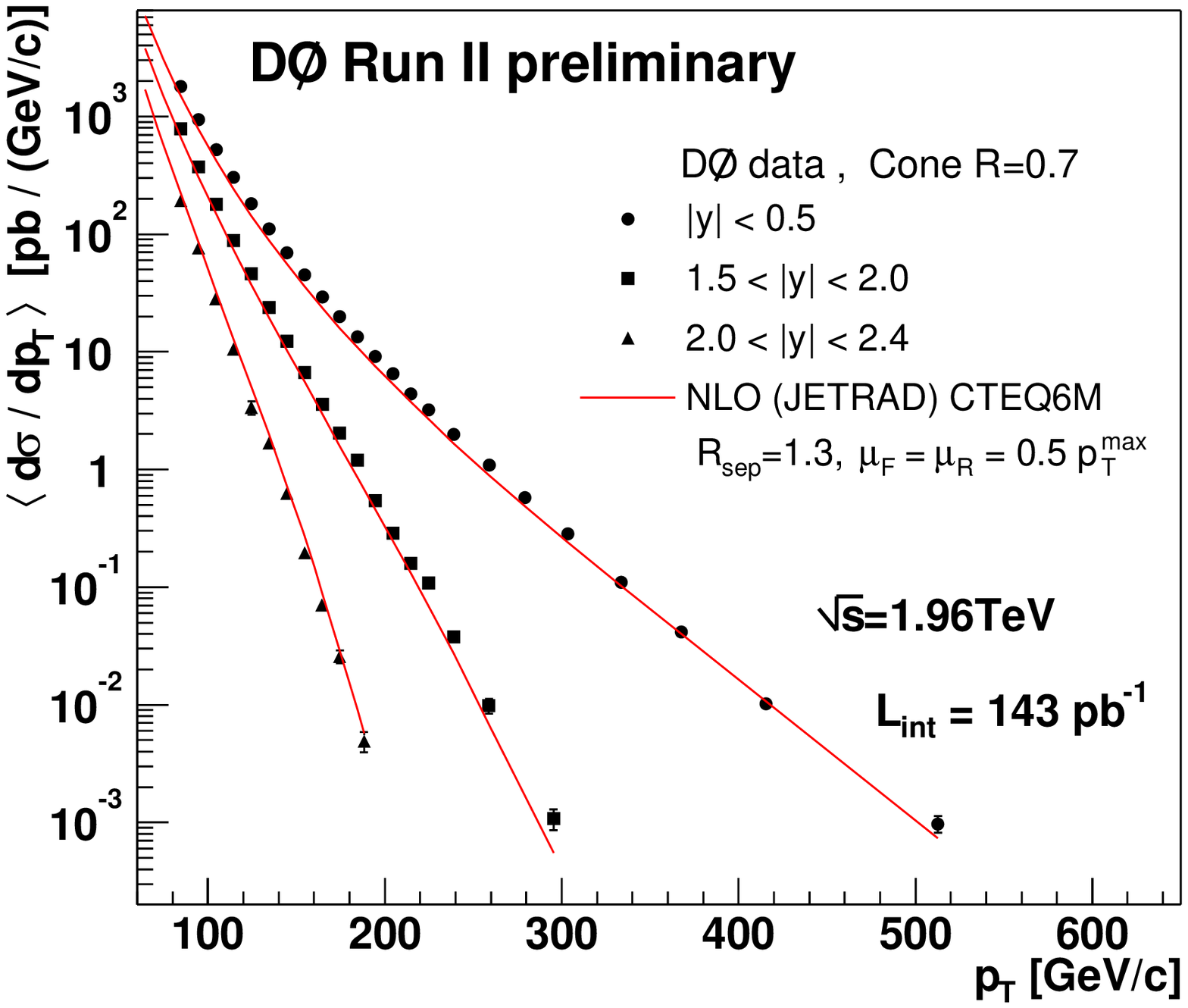}{}}
\vskip 1.5cm
\hspace*{0.cm} a)
\end{center}

\vskip 5cm

\begin{center}
\hspace*{-2.5cm}
\parbox{5cm}{\epsfxsize=5.cm\epsfysize=5.cm\epsfbox[95 95 400 400]
{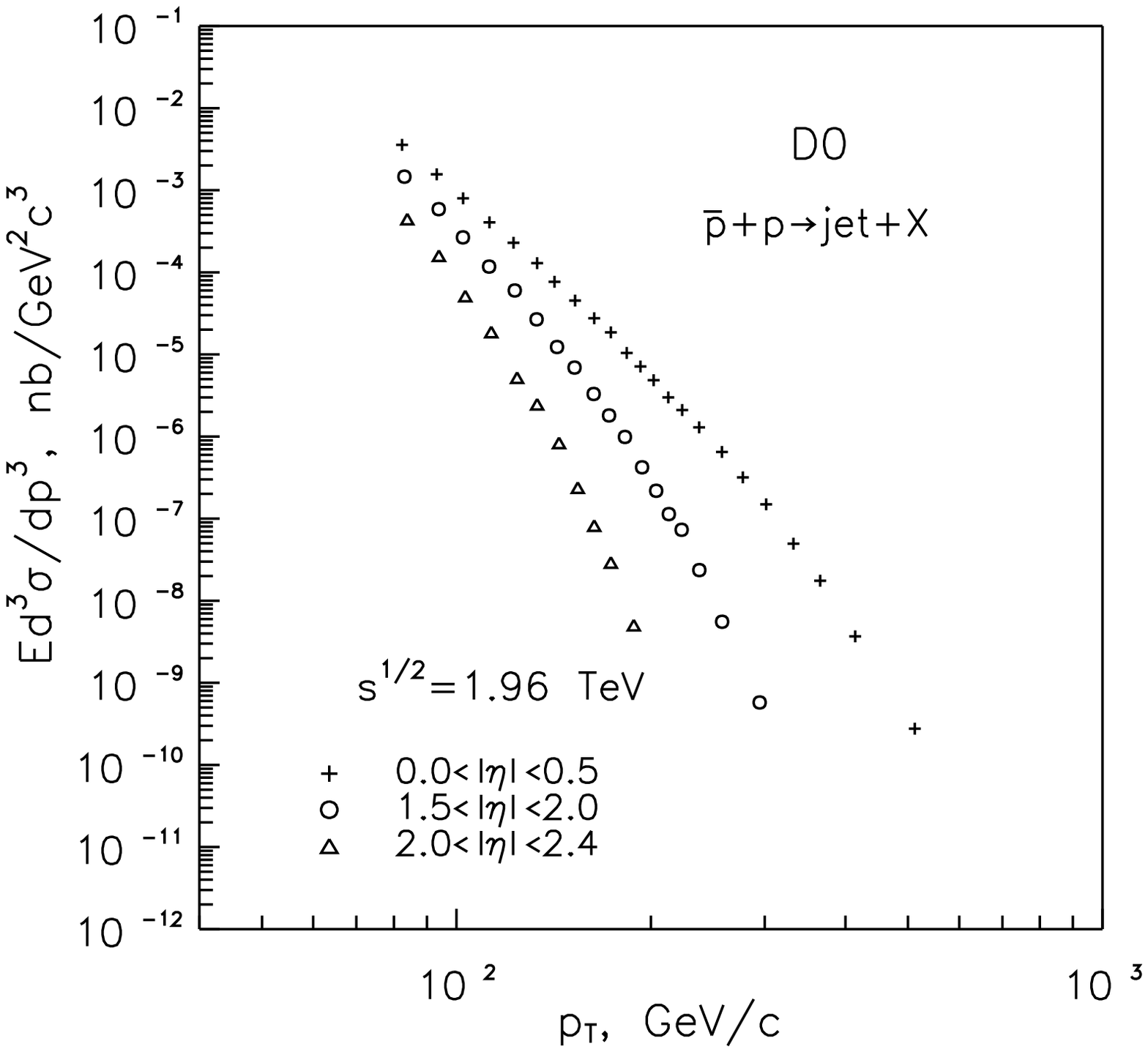}{}}
\hspace*{3cm}
\parbox{5cm}{\epsfxsize=5.cm\epsfysize=5.cm\epsfbox[95 95 400 400]
{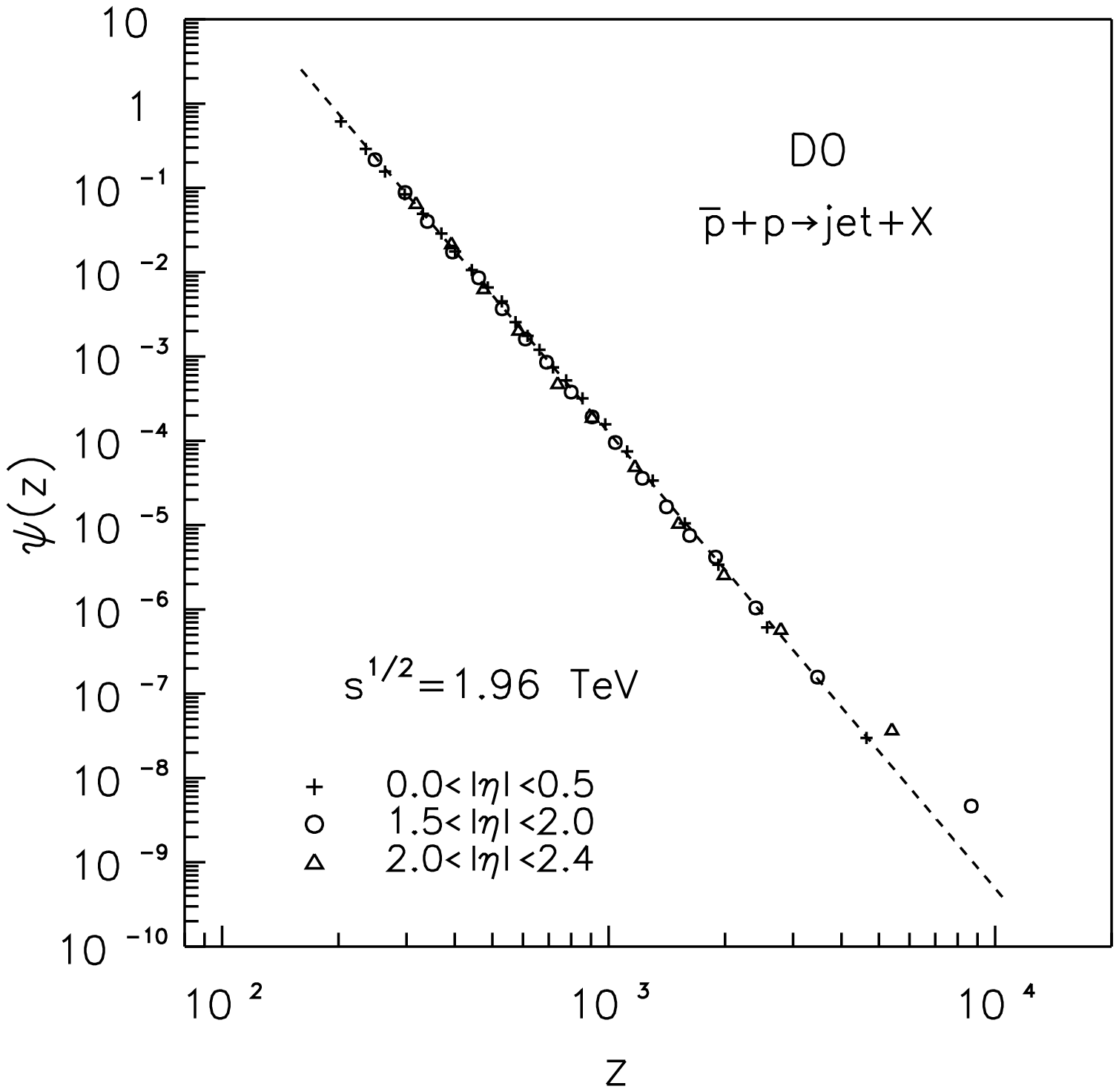}{}}
\vskip -1.cm
\hspace*{0.cm} b)\hspace*{8.cm} c)\\[0.5cm]
\end{center}

{\bf Figure 4.}
a) The preliminary D0 data  \cite{Chleb} on inclusive spectrum  $d\sigma /dp_T$
of one jets produced in $\bar p-p $ collisions at $\sqrt s = 1960$~GeV
in the different pseudorapidity ranges  $0.<|\eta|<0.5$, $2.5<|\eta|<2.0$
 and  $2.0<|\eta|<2.4$ as a function of the transverse momentum $p_T$.
(b) The D0 data  on invariant cross section $Ed^3\sigma/dp^3$
of jet production at $\sqrt s = 1960$~GeV \cite{Chleb}
in $p_T$- and (c) $z$-presentations.
The dashed line represents the power fit to the data.
\vskip 0.5cm

\newpage
\begin{minipage}{4cm}

\end{minipage}

\vskip 0cm
\begin{center}
\hspace*{-1.5cm}
\parbox{6.cm}{\epsfxsize=6.cm\epsfysize=6.cm\epsfbox[95 95 400 400]
{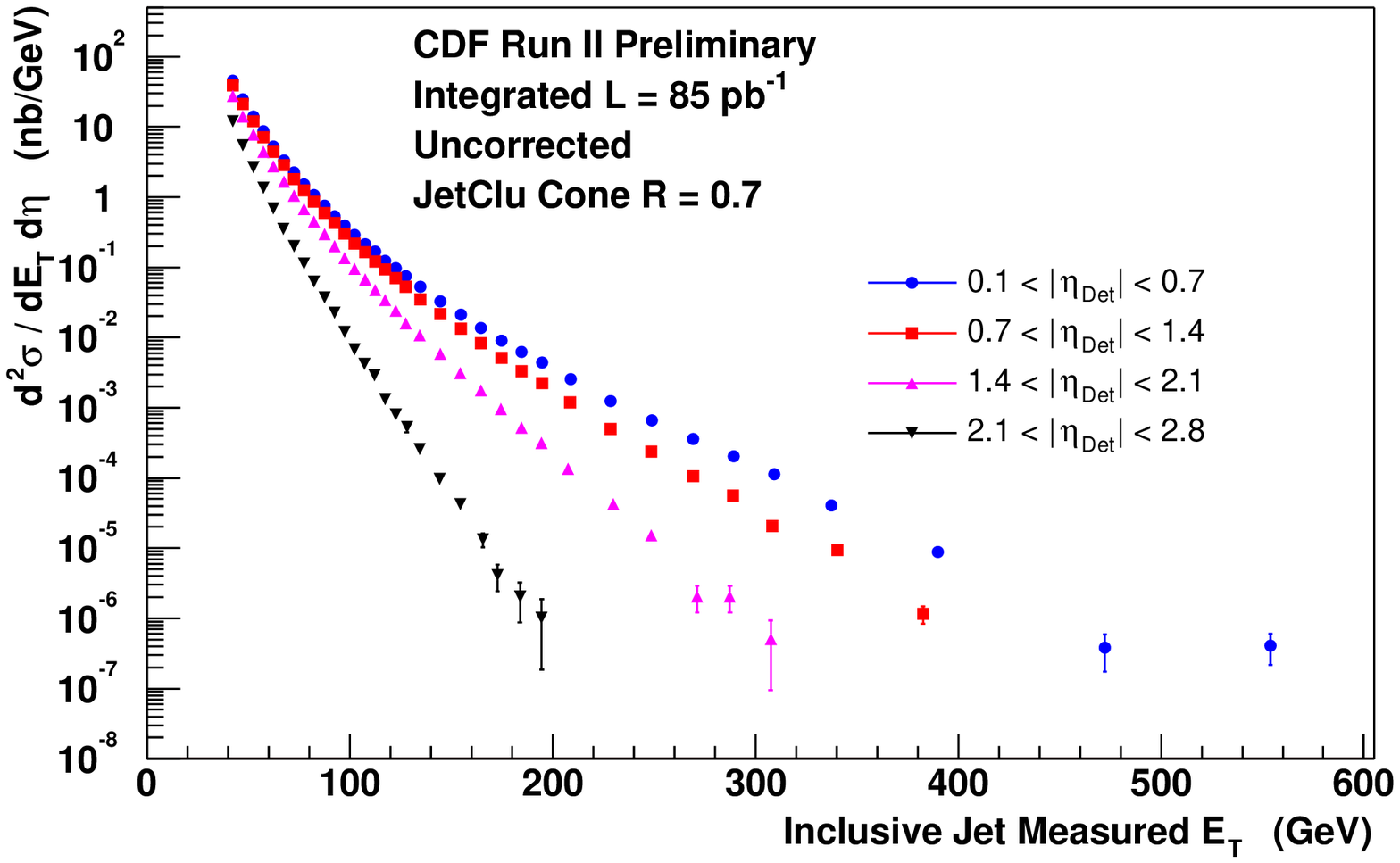}{}}
\vskip  2.5cm
\hspace*{0.cm} a)
\end{center}

\vskip 5cm

\begin{center}
\hspace*{-2.5cm}
\parbox{5cm}{\epsfxsize=5.cm\epsfysize=5.cm\epsfbox[95 95 400 400]
{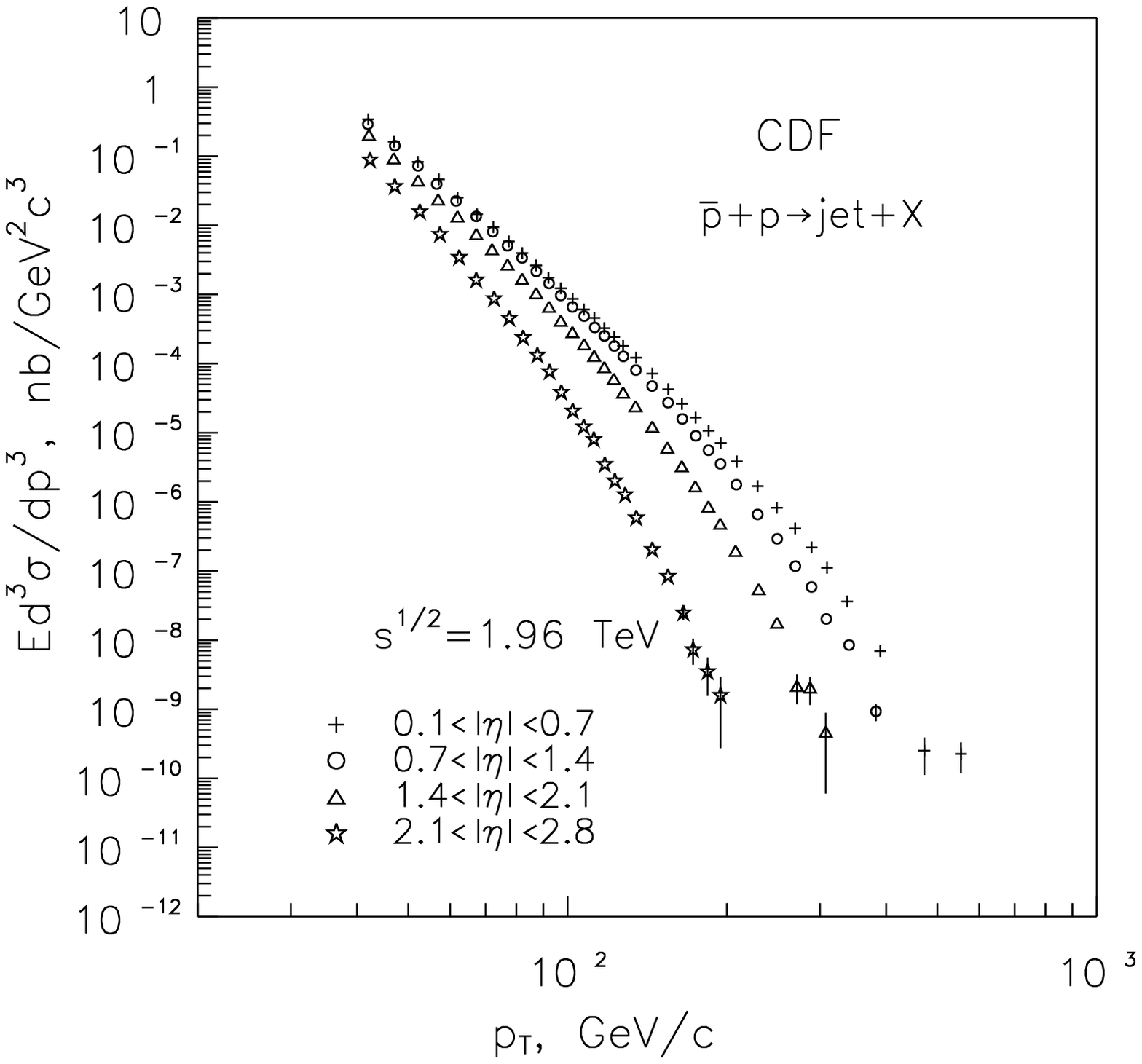}{}}
\hspace*{3cm}
\parbox{5cm}{\epsfxsize=5.cm\epsfysize=5.cm\epsfbox[95 95 400 400]
{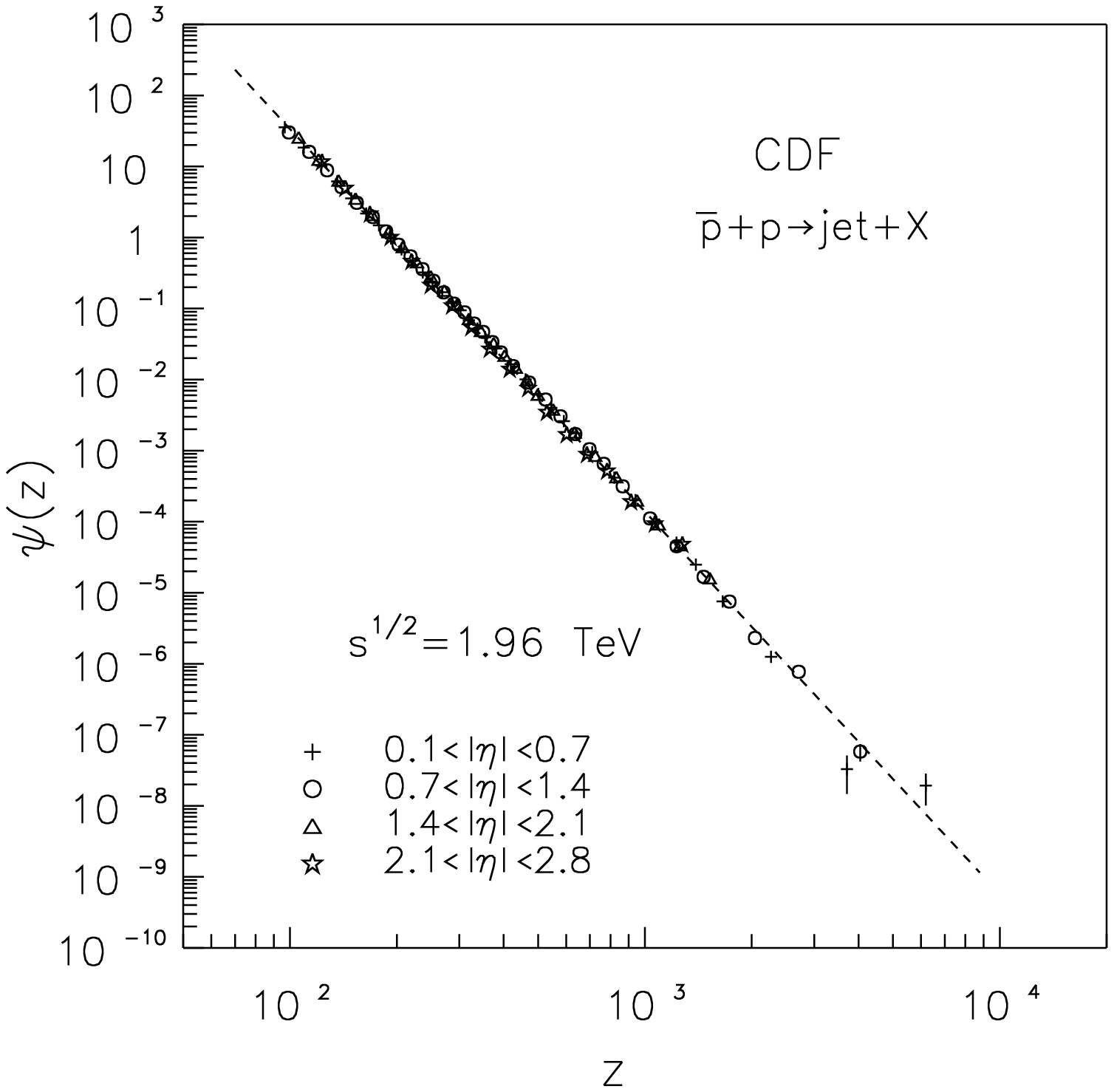}{}}
\vskip -1.cm
\hspace*{0.cm} b)\hspace*{8.cm} c)\\[0.5cm]
\end{center}

{\bf Figure 5.}
a) The preliminary CDF data  \cite{Tonnes} on inclusive spectrum
 $d^2\sigma/dE_Td\eta$ of one jets produced in $\bar p-p $ collisions
 at $\sqrt s = 1960$~GeV
in the different pseudorapidity ranges  $0.1<|\eta|<0.7$, $0.7<|\eta|<1.4$,
$1.4<|\eta|<2.1$, and  $2.1<|\eta|<2.8$
as a function of the transverse momentum $p_T$.
(b) The CDF data  on invariant cross section $Ed^3\sigma/dp^3$
of jet production at $\sqrt s = 1960$~GeV \cite{Tonnes}
in $p_T$- and (c) $z$-presentations.
The dashed line represents the power fit to the data.
\vskip 0.5cm

\newpage
\begin{minipage}{4cm}

\end{minipage}

\vskip 6cm
\begin{center}
\hspace*{-2.5cm}
\parbox{7.cm}{\epsfxsize=7.cm\epsfysize=7.cm\epsfbox[95 95 400 400]
{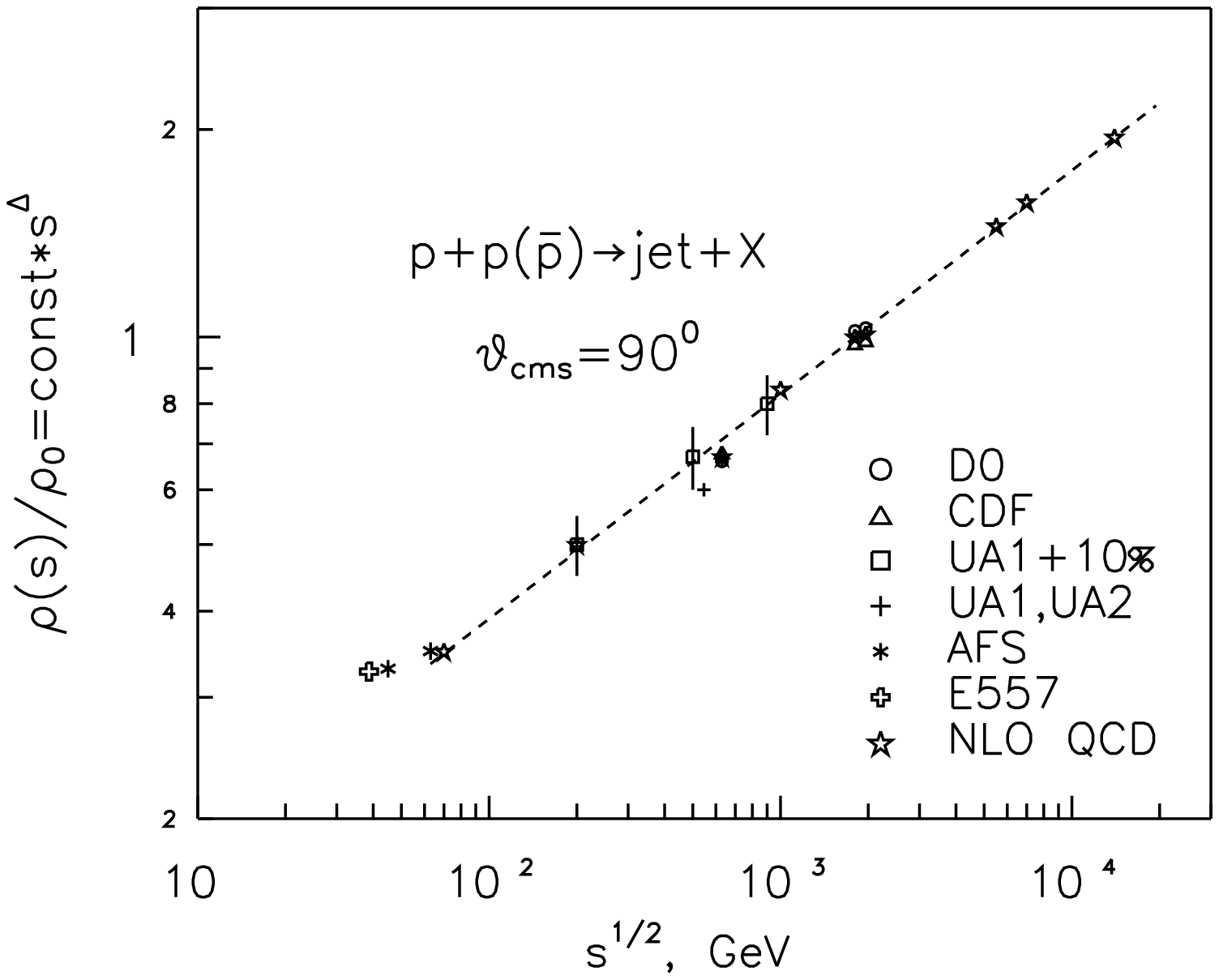}{}}
\end{center}


{\bf Figure 6.}
The dependence of the normalized jet multiplicity density $\rho(s)/\rho_0$
in $\bar p-p $ collisions on energy $\sqrt s$.
The dashed line represents the power fit to the data.

\end{document}